\documentclass[aps,prl,twocolumn,reprint,superscriptaddress]{revtex4}

\usepackage{hyperref}
\usepackage{graphicx}  
\usepackage{bm}        
\usepackage{amssymb}   
\usepackage{xspace}
\usepackage{verbatim}
\usepackage{color}
\usepackage{soul}
\usepackage{subfigure}
\usepackage[export]{adjustbox}
\usepackage{dcolumn}
\usepackage{amsthm,stmaryrd,mathtools}
\usepackage{txfonts}
\usepackage{braket}
\usepackage{amsmath}
\usepackage{xcolor}
\usepackage{array}
\usepackage{multirow}
\usepackage{tabularx}
\usepackage{booktabs} 
\usepackage{lipsum}
\usepackage{dsfont}

\begin{document}

\author{Rozhin Yousefjani}%
\email{ryousefjani@hbku.edu.qa}
\affiliation{Institute of Fundamental and Frontier Sciences, University of Electronic Science and Technology of China, Chengdu 610051, China}
\affiliation{Qatar Center for Quantum Computing, College of Science and Engineering, Hamad Bin Khalifa University, Doha, Qatar}

\author{Angelo Carollo}%
\affiliation{Dipartimento di Fisica e Chimica “E. Segr\`{e}”, Group of Interdisciplinary Theoretical Physics, Universit`a degli Studi di Palermo, I-90128 Palermo, Italy}

\author{Krzysztof Sacha}%
\affiliation{Instytut Fizyki Teoretycznej, Wydział Fizyki, Astronomii i Informatyki Stosowanej,
Uniwersytet Jagiello\'{n}ski, ulica Profesora Stanisława Łojasiewicza 11, PL-30-348 Krak\'{o}w, Poland, Centrum Marka Kaca, Uniwersytet Jagiello\'{n}ski,
ulica Profesora Stanisława Łojasiewicza 11, PL-30-348 Krak\'{o}w, Poland}

\author{Saif Al-Kuwari}%
\affiliation{Qatar Center for Quantum Computing, College of Science and Engineering, Hamad Bin Khalifa University, Doha, Qatar}

\author{Abolfazl Bayat}%
\affiliation{Institute of Fundamental and Frontier Sciences, University of Electronic Science and Technology of China, Chengdu 610051, China}
\affiliation{Key Laboratory of Quantum Physics and Photonic Quantum Information, Ministry of Education, University of Electronic Science
and Technology of China, Chengdu 611731, China}

\title{Non-Hermitian Discrete Time Crystals}

\begin{abstract}
{Discrete time crystals (DTC) exhibit a special non-equilibrium phase of matter in periodically driven many-body systems with spontaneous breaking of time translational symmetry. 
The presence of decoherence generally enhances thermalization and destroys the coherence required for the existence of DTC.
In this letter, we devise a mechanism for establishing a stable DTC with period-doubling oscillations in an open quantum system that is governed by a properly tailored non-Hermitian Hamiltonian. We find a specific class of non-reciprocal couplings in our non-Hermitian dynamics which prevents thermalization through  eigenstate ordering. Such choice of non-Hermitian dynamics, significantly enhances the stability of the DTC against imperfect pulses. Through a comprehensive analysis, we determine the phase diagram of the system in terms of pulse imperfection.
}     
\end{abstract}

\maketitle

{\it Introduction.--}
Drawing an analogy to conventional crystals, which emerge from the breaking of spatial translational symmetries, the concept of spontaneously breaking time translation symmetry has sparked a rapidly increasing interest in the formation of time crystals~\cite{wilczek2012quantum}. However, it has been shown that time crystals cannot emerge in equilibrium in systems with two-body interactions~\cite{Bruno2013b,watanabe2015absence}, see also \cite{Kozin2019}.
Instead, externally driven non-equilibrium systems have been proposed~\cite{Sacha2015,khemani2016phase,else2016floquet,sacha2017time} and then experimentally verified~\cite{zhang2017observation,choi2017observation,pal2018temporal,rovny2018observation,smits2018observation,autti2018observation,Randall2021,Kessler2021,Kyprianidis2021,xu2021realizing,Taheri2022,mi2022time,frey2022realization,Liu2023,he2024experimental,Bao2024,Kazuya2024,Liu2024a,Liu2024} as a testbed for breaking discrete time translational symmetry and the formation of Discrete Time Crystals (DTC). In general, in a periodically driven system with period $T$, a DTC is identified through several features: (i) the observables oscillate with period $gT$, where $g{>}1$ is an integer number; (ii) the oscillations of the system are robust against imperfections in the driving pulse; and (iii) in the thermodynamic limit, thermalization is prevented and oscillations persist indefinitely~\cite{sacha2017time,else2020discrete,khemani2019brief,SachaTC2020,Zaletel2023}.
In closed systems, where evolution is unitary, DTCs have been engineered with the aid of, e.g., many-body localization~\cite{khemani2016phase,else2016floquet,yao2017discrete,ho2017critical,khemani2019brief,else2020discrete,Iemini2023}, 
bosonic self-trapping \cite{Wuster2012,Sacha2015,russomanno2017floquet,Giergiel2018c,Matus2019,pizzi2021higher,SachaTC2020}, 
Stark gradient fields \cite{liu2023discrete,kshetrimayum2020stark}, gradient interaction \cite{yousefjani2024DTC}, 
domain-wall confinement~\cite{Collura2022Discrete}, 
Floquet integrability~\cite{huang2018clean} 
and quantum scars \cite{Maskara2021Discrete,Deng2023Using,Bull2022Tuning,Huang2022Discrete,Huang2023Analytical}. 
DTCs in closed systems are usually fragile if they are in contact with the environment \cite{breuer2002theory,lazarides2017fate,riera2020time}. However, dissipation can also be an ally if it is part of the mechanism responsible for the spontaneous emergence of stable periodic evolution. In so-called dissipative DTCs, dissipation serves as a channel for releasing excess energy pumped by a periodic external disturbance \cite{Gong2018,gambetta2019discrete,lazarides2020time}. This type of time crystal has been demonstrated in the laboratory 
 \cite{Kessler2021,Taheri2022} and represents one branch of research.

Non-Hermitian Hamiltonians naturally arise as a well-established framework for the effective description of open system dynamics~\cite{Rotter2009,plenio1998quantum,daley2014quantum,Bender2003,Bender2007}.
Within this class of Hamiltonians, a special subset is the non-reciprocal systems, i.e. Hamiltonians describing asymmetric strength interactions between some of their constituents. Non-reciprocal Hamiltonians emerge in various forms throughout physics (see, e.g. Ref.~\cite{Fruchart2021} and references therein), and have garnered considerable attention due to the unique phenomena they exhibit. These phenomena include, for instance, the extreme sensitivity of spectral properties to boundary conditions~\cite{Hatano1996}, the emergence of novel topological characteristics in the complex spectrum~\cite{ashida2020non,Bergholtz2021,kawabata2019symmetry}, directional amplification and transport~\cite{schomerus2020nonreciprocal}, violation of bulk-boundary correspondence~\cite{zhang2022universal,li2020critical,xiao2020non}, and continuous classical time crystals \cite{Raskatla2024}.
Apart from fundamental interests, non-reciprocal systems have applications in quantum sensing~\cite{wiersig2014enhancing,montenegro2024review,mcdonald2020exponentially,chen2017exceptional,chen2018generalized,zhao2018exceptional,hodaei2017enhanced,sarkar2024critical,budich2020non,koch2022quantum}, quantum thermodynamics~\cite{ahmadi2024nonreciprocal}, and quantum control~\cite{metelmann2015nonreciprocal,shen2016experimental}. 
A natural and largely open question is whether one can benefit from open quantum systems, described by non-Hermitian dynamics, for developing a DTC that is stable against decoherence and robust against imperfect pulses.
\begin{figure}
    \centering
    \includegraphics[width=0.51\linewidth]{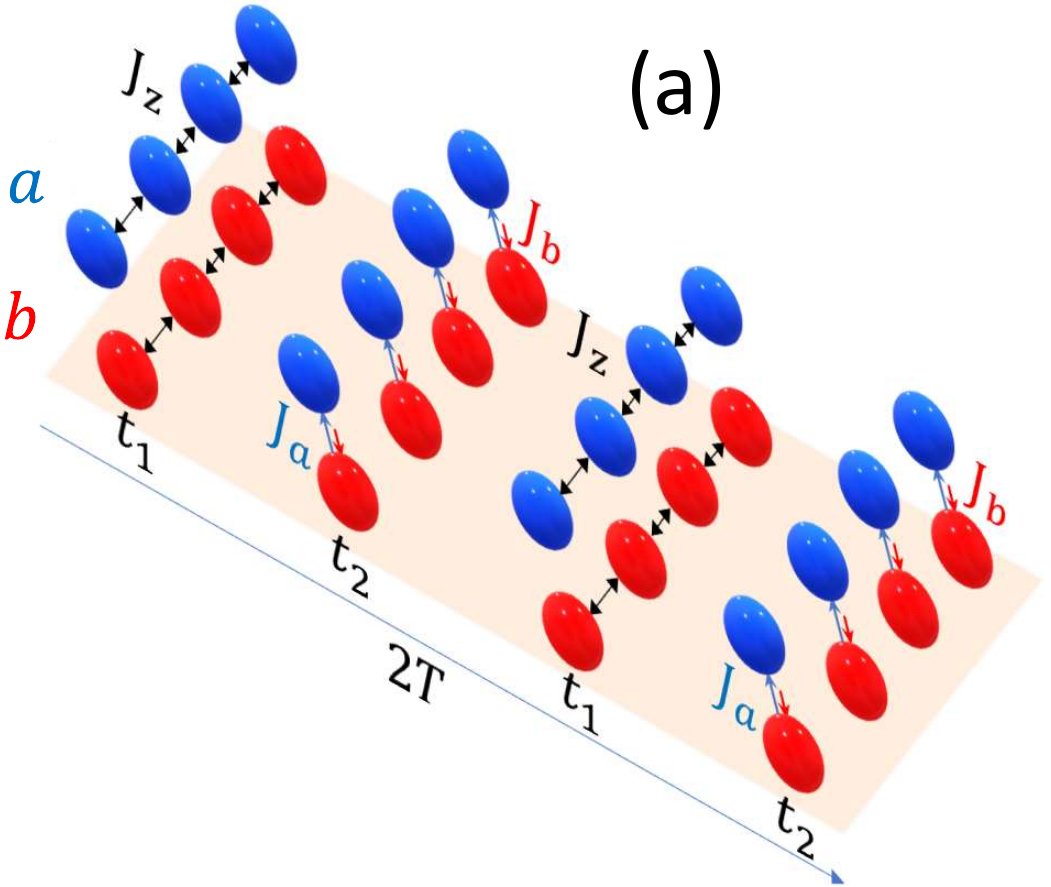}
 \includegraphics[width=0.47\linewidth]{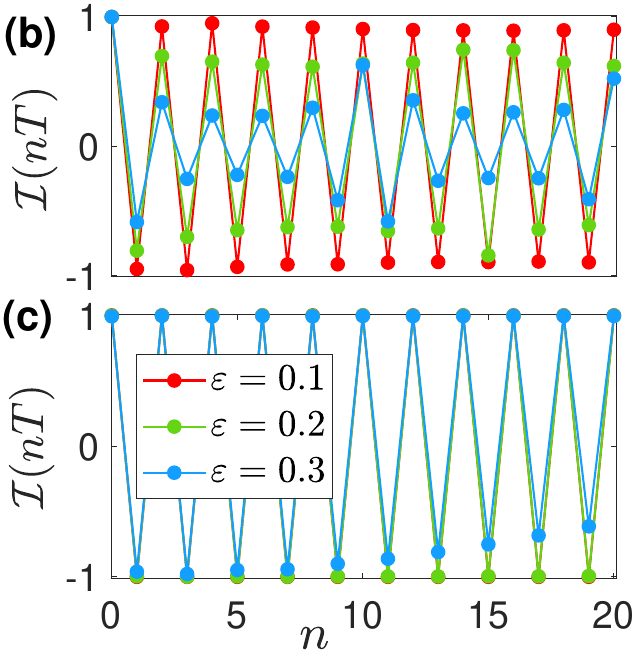} 
    \caption{(a) Schematic for the DTC model used here. Imbalance in the total magnetization of the chains $a$ and $b$ denoted as $\mathcal{I}(nT)$, versus  period cycling $n$ in (b) H-DTC, namely $(\varepsilon_a,\varepsilon_b){=}(\varepsilon,\varepsilon)$, and (c) NH-DTC, namely 
$(\varepsilon_a,\varepsilon_b){=}(\varepsilon,{-}\varepsilon)$, in a chain of size $L{=}8$. }
    \label{fig:Fig1}
\end{figure}

In this letter, we address the above open questions by proposing a mechanism for establishing a DTC with period-doubling oscillations in an integrable model governed by a non-Hermitian Hamiltonian. We show that by properly engineering the non-reciprocal couplings in non-Hermitian dynamics one can prevent thermalization through inducing eigenstate ordering which significantly enhances the robustness of the DTC against pulse imperfections. 
\\

{\it Model.--} 
We consider two one-dimensional chains $a$ and $b$, each containing $L$ spin-$1/2$ particles interacting via the following periodic Hamiltonian
\begin{equation}\label{Eq.Hamiltonian}
H(t) = 
\begin{cases}
H_a+H_b=-J_z\sum_{\mu=a}^b\sum_{j=1}^{L-1}\sigma_j^{\mu z}\sigma_{j+1}^{\mu z} & 0<t\leqslant t_1 \\
H_{I} = \sum_{j=1}^{L} (J_a \sigma_j^{a+}\sigma_{j}^{b-} + J_b \sigma_j^{a-}\sigma_{j}^{b+}) & t_1<t\leqslant t_1 + t_2
\end{cases}  
\end{equation}
with $J_z$ being spin exchange coupling in each chain $\mu{=}a,b$, $\sigma^{\mu\pm}_{j}{=}\frac{1}{2}(\sigma^{\mu x}_{j} {\pm} i\sigma^{\mu y}_{j})$, and $\sigma^{\mu(x,y,z)}_{j}$ representing
Pauli operators acting at site $j$ of chain $\mu$. Note that $H_I$ is generally non-Hermitian and represents directional-dependent hopping of the spin excitations between the two chains. Fig.~\ref{fig:Fig1} (a) shows a schematic of the system. We consider $J_{\mu}{=}\frac{J_z\pi}{2}(1{+}\varepsilon_{\mu})$, where dimensionless parameters $\varepsilon_{\mu}{\in}\mathds{R}$ quantify deviation from a perfect coupling of $\frac{J_z\pi}{2}$ in each chain. Therefore, the dynamics of the system is determined by $(\varepsilon_a,\varepsilon_b)$, such that by choice of $\varepsilon_a{=}\varepsilon_b$, the system becomes Hermitian and any other choices result in non-Hermitian, non-reciprocal dynamics. 
The interaction between particles is repeated with period $T{=}t_1{+}t_2$ and we set $t_1{=}t_2{=}\frac{1}{2J_z}$. 
Considering $J_z$ as the unit of energy, the single-cycle Floquet operator is expressed as 
\begin{equation}\label{eq:Floquet_Operator}
 U_F(\varepsilon_a,\varepsilon_b) = e^{\frac{-i}{2J_z} H_I} e^{\frac{-i}{2J_z} (H_a+H_b)}. 
\end{equation}
In the following, we show how the above mechanism leads to the emergence of period-doubling DTC and how different choices of $(\varepsilon_a,\varepsilon_b)$ play a role in stabilizing the DTC.  
We analyze two specific cases, namely, Hermitian DTC (H-DTC) obtained for $\varepsilon_{a}{=}\varepsilon_b$, and non-Hermitian DTC (NH-DTC) achieved for $\varepsilon_a{\neq}\varepsilon_b$.
The former leads to a symmetric hopping rate between chains $a$ and $b$, thereby a reciprocal dynamic and a unitary Floquet operator $U_F$. On the other hand, the latter asymmetric hopping generates a non-reciprocal dynamic and a non-unitary Floquet operator.
While increasing the deviation $(\varepsilon_a,\varepsilon_b)$ is in general detrimental to the DTC ordering, we show that there is a class of such choices, given by $(\varepsilon_a,\varepsilon_b){=}(\varepsilon,-\varepsilon)$, which enhance the stability of DTC. \\

{\it Discrete time crystal.--}
To observe the time-crystalline temporal order, one can first initialize the chains $a$ and $b$ in a fully polarized state but with opposite total magnetization in the $z$-direction, namely $|\psi(0)\rangle{=}|{\uparrow,\cdots,\uparrow}\rangle_{a}{\otimes}|{\downarrow,\cdots,\downarrow}\rangle_{b}$.
Then measure the imbalance in the total magnetization of chains during stroboscopic times, defined as
\begin{align}
    \mathcal{I}(nT)&= \frac{1}{L} \sum_{j=1}^{L}\mathcal{I}_{j}(nT)  
\end{align}
in which $\mathcal{I}_{j}(nT){=}\frac{1}{2}\langle \psi(nT)|(\sigma_{j}^{az}{-}\sigma_{j}^{bz})|\psi(nT)\rangle$. 
In the ideal case, namely in the absence of deviation $(\varepsilon_a,\varepsilon_b){=}(0,0)$, dynamics under $H_{I}$ leads to coherent oscillations of the system between $|\psi(0)\rangle$ and its global spin-inversion  partner  $|\tilde{\psi}(0)\rangle{=}|{\downarrow,{\cdots},\downarrow}\rangle_{a}{\otimes}|{\uparrow,{\cdots},\uparrow}\rangle_{b}$.
Consequently, one obtains 
$\mathcal{I}(nT){=}1$, and ${-}1$, in even  and odd $n$'s, respectively. 
Restoring the system to its initial configuration after $2T$, i.e. period doubling, verifies that the oscillation of $\mathcal{I}(nT)$ is synchronized at half drive frequency, a phenomenon known as subharmonic locking~\cite{tiana2022time}.
For imperfect coherent pulses with $(\varepsilon_a,\varepsilon_b){=}(\varepsilon,\varepsilon)$, the oscillation of $\mathcal{I}(nT)$ is plotted in Fig.~\ref{fig:Fig1}(b) for various choices of $\varepsilon$ which shows that the H-DTC preserves periodic doubling and is moderately stable against imperfections up to $\varepsilon{\simeq}0.2$. For imperfect incoherent pulses with $(\varepsilon_a,\varepsilon_b){=}(\varepsilon,-\varepsilon)$, the imbalance is plotted in Fig.~\ref{fig:Fig1}(c) for various choices of $\varepsilon$ which shows that period-doubling survives in such a NH-DTC with significantly enhanced stability, even up to $\varepsilon{\simeq}0.3$. 
As Fig.~\ref{fig:Fig1}(c) clearly shows, the nonreciprocal coupling between the chains $a$ and $b$ caused by setting $(\varepsilon_a,\varepsilon_b){=}(\varepsilon,-\varepsilon)$  
results in an asymmetric quantum state evolution which is very pronounced for $\varepsilon{=}0.3$. 

\begin{figure}[t]
    \centering
    \includegraphics[width=\linewidth]{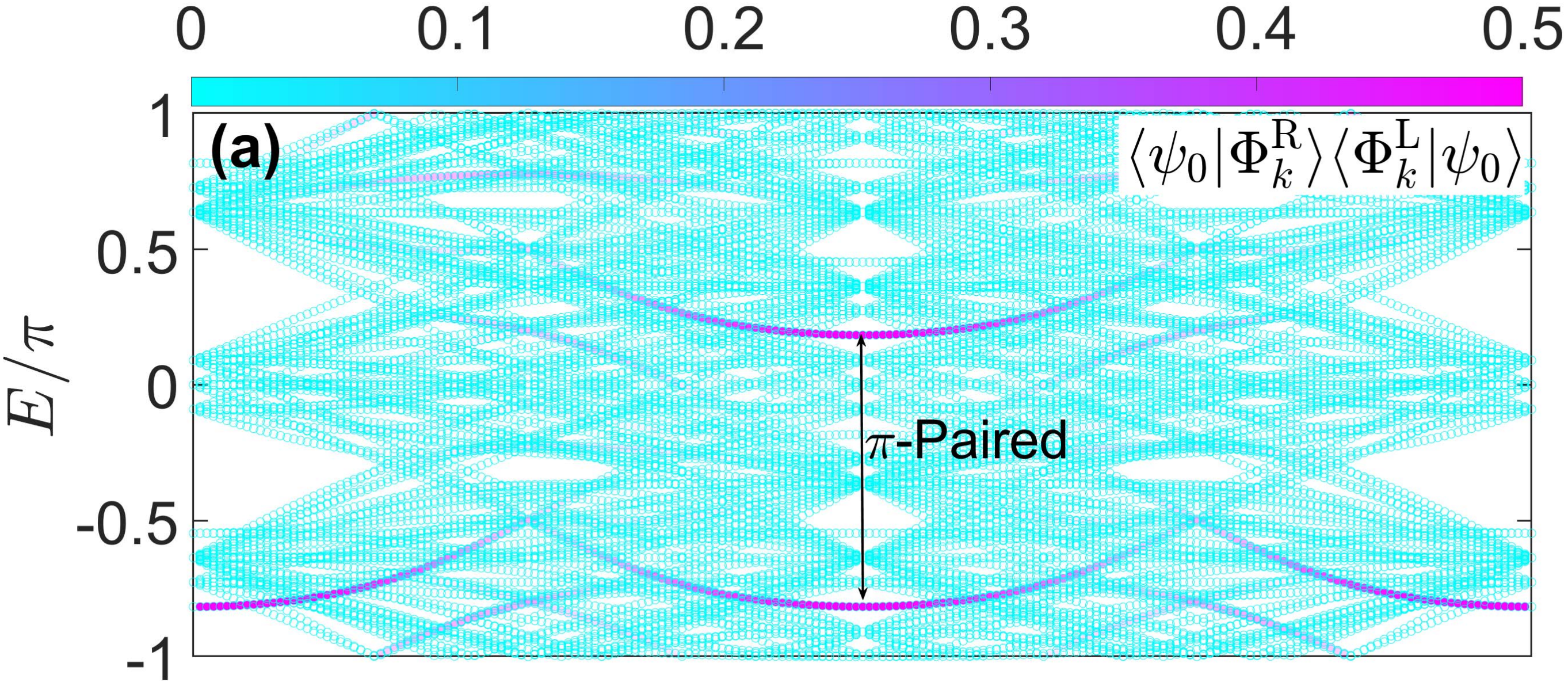}
    \includegraphics[width=\linewidth]{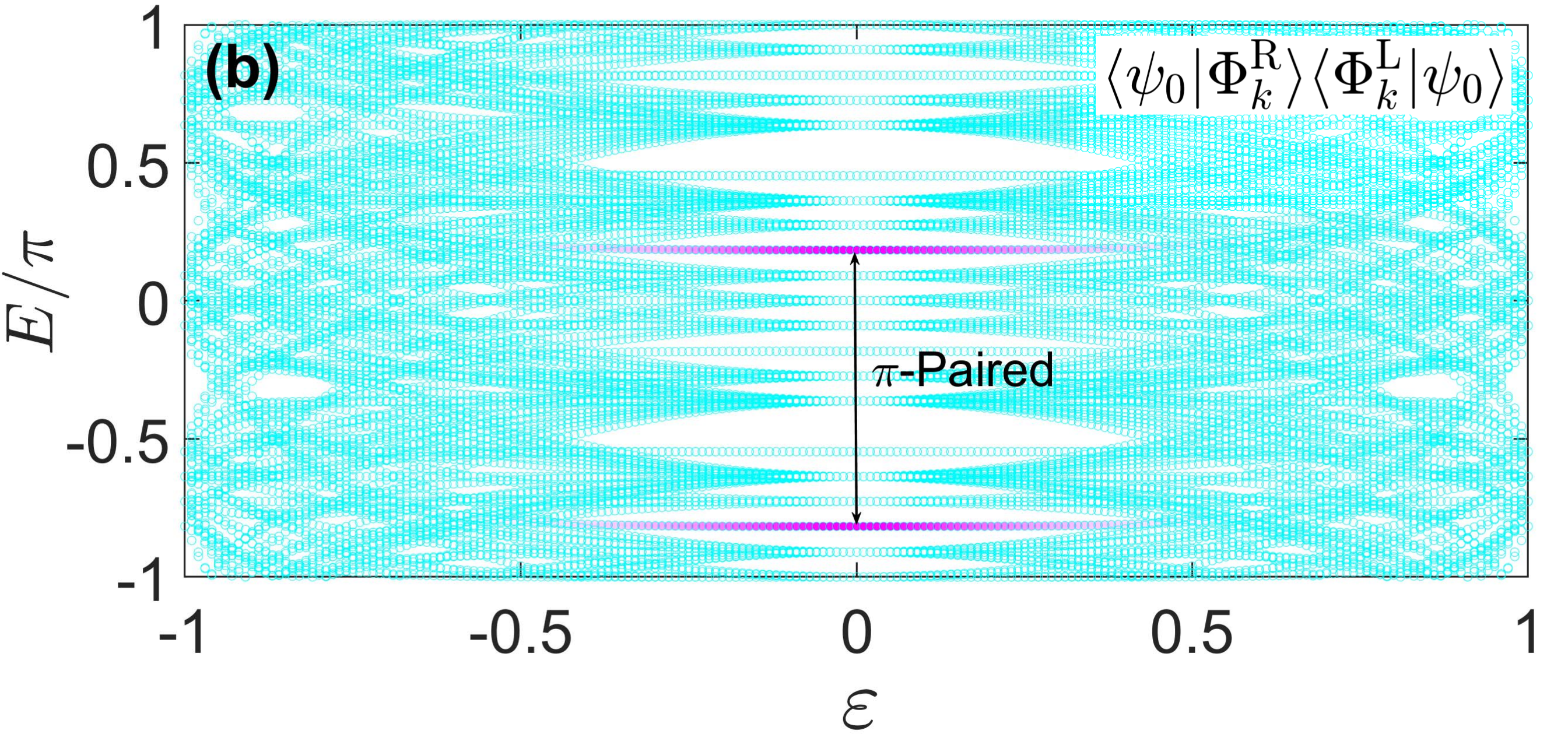}
    \caption{The overlap of the Floquet operator's spectrum $\Big(U_{F}(\varepsilon_a,\varepsilon_b){=}\sum_{k}e^{-iE_{k}}|\Phi_{k}^{\rm R}\rangle \langle\Phi_{k}^{\rm L}|\Big)$ with $|\psi(0)\rangle$ as a function of the quasienergy $E$ and the imperfection $\varepsilon$, in (a) H-DTC $(\varepsilon_a,\varepsilon_b){=}(\varepsilon,\varepsilon)$ and (b) NH-DTC $(\varepsilon_a,\varepsilon_b){=}(\varepsilon,-\varepsilon)$ with size $L{=}6$. }
    \label{fig:Spectrum}
\end{figure}

To clarify the origin of the period-doubling oscillations, we analyze the eigenstates of the Floquet operator of Eq.~(\ref{eq:Floquet_Operator}) and its dimensionless quasienergies.
For the sake of generality, we use formal algebra for non-Hermitian physics and then discuss the reduction to the Hermitian case.  
Floquet states of $U_F(\varepsilon_a,\varepsilon_b){=}e^{-iH_{\mathrm{eff}}(\varepsilon_a,\varepsilon_b)}$ 
with $H_{\mathrm{eff}}$ as an effective Hamiltonian are defined as $U_F(\varepsilon_a,\varepsilon_b){=}\sum_{k}e^{-iE_{k}}|\Phi_k^{\rm{R}}\rangle\langle \Phi_k^{\rm{L}}|$.  
Here $|\Phi_k^{\rm{R,L}}\rangle{=}|\phi_k^{\rm{R,L}}\rangle{/}\sqrt{|\langle \phi_k^{\rm{L}} | \phi_k^{\rm{R}} \rangle|}$ are the normalized right and left eigenstates of the $H_{\mathrm{eff}}$ with
$H_{\mathrm{eff}}|\phi_k^{\rm{R}}\rangle{=}E_{k}|\phi_k^{\rm{R}}\rangle$, and $H_{\mathrm{eff}}^{\dagger}|\phi_k^{\rm{L}}\rangle{=} E_{k}^{*}|\phi_k^{\rm{L}}\rangle$.
Note that the Floquet states are biorthogonal, namely upon some normalization give $\langle \Phi_l^{\rm{L}} | \Phi_k^{\rm{R}} \rangle{=}\delta_{lk}$ with the completeness relation as $\sum_{k}|\Phi_k^{\rm{R}}\rangle\langle \Phi_k^{\rm{L}}|{=}\mathds{1}$~\cite{brody2013biorthogonal}.
For $\varepsilon_{a}{=}\varepsilon_{b}$ and, hence, Hermitian $H_{\mathrm{eff}}$ one expects $E_{k}{=}E_{k}^*$ and $|\Phi_k^{\rm{R}}\rangle{=}|\Phi_k^{\rm{L}}\rangle$.
Although, in general, non-Hermitian matrices are expected to have complex eigenvalues, those non-Hermitian Hamiltonians that do not break the PT-symmetry support real energy spectrum, namely $E_{k}{=}E_{k}^*$~\cite{krasnok2021PTS,mochizuki2016explicit}.
Remarkably, as we will discuss in the next section, the Hamiltonian $H_{\rm eff}$ is PT-symmetric, and for all values of $(\varepsilon_a,\varepsilon_b)$ the system does not break the PT symmetry, which implies that all quasienergies are real.

The energy spectrum of period-doubling DTCs, exhibits a notable order, namely "Schr{\"o}dinger cat" eigenstates with eigenvalues differing by $\pi$ are consistently paired, known as $\pi$-Paired eigenstate~\cite{khemani2016phase,russomanno2017floquet,estarellas2020simulating}.
Here, for small imperfections, i.e. $|\varepsilon_a|,|\varepsilon_b|{\ll} 1$, one can see that the spectrum of $U_{F}(\varepsilon_a,\varepsilon_b)$ contains eigenstates with quasienergies separated by $\pi$ that strongly overlap with $|\psi(0)\rangle$ and $|\tilde{\psi}(0)\rangle$. 
This can be seen in 
Fig.~\ref{fig:Spectrum}, where we depict $\langle \psi(0)|\Phi_k^{\rm{R}}\rangle\langle\Phi_k^{\rm{L}}|\psi(0)\rangle$ as a function of quasienergy $E$ and imperfection strength $\varepsilon$ for H-DTC, namely $(\varepsilon,\varepsilon)$, in panel (a) and NH-DTC, namely $(\varepsilon,\varepsilon)$, in (b).
The results for $\langle \tilde{\psi}(0)|\Phi_k^{\rm{R}}\rangle\langle\Phi_k^{\rm{L}}|\tilde{\psi}(0)\rangle$ in both cases are identical with Fig.~\ref{fig:Spectrum} (data has not been shown).
This implies that the eigenstates can be well approximated by the long-range correlated ``Schr{\"o}dinger cat” states as $|\psi_{\pm}\rangle{=}(|\psi(0)\rangle {\pm}|\tilde{\psi}(0)\rangle){/}\sqrt{2}$.
In both cases, by increasing the imperfection strength $|\varepsilon|$, the overlap of the initial state with the $\pi$-Paired eigenstates decreases, signaling dispersion through contribution of other eigenstates and thus losing  temporal order.
Interestingly, the eigenstate order in NH-DTC shows more resistance against imperfection than the one in H-DTC.  \\
\begin{figure}[t]
    \centering
    \includegraphics[width=0.49\linewidth]{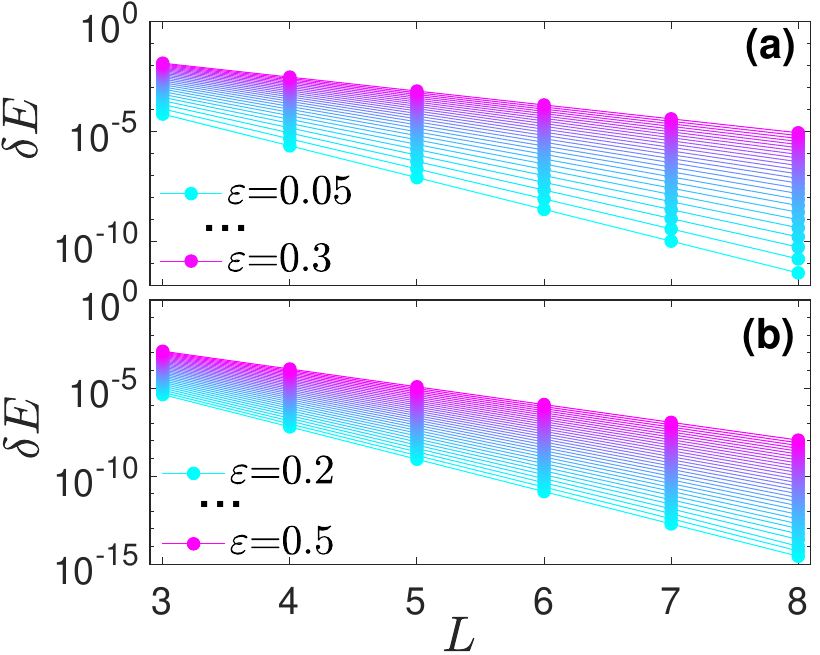}
    \includegraphics[width=0.49\linewidth]{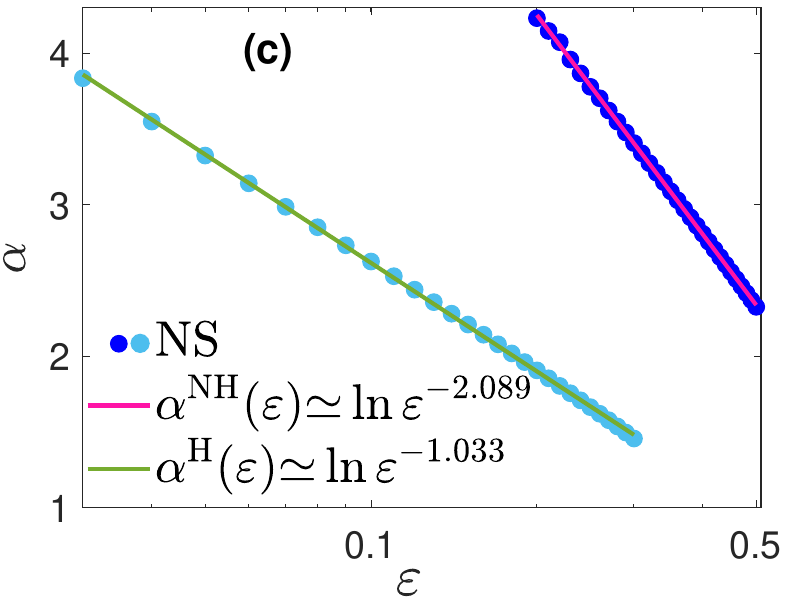}
    \caption{Deviation of the quasienergy gap $\delta E$ as a function of system size $L$ for different values of $\varepsilon$ in (a) H-DTC and (b) NH-DTC. Fitting function $\delta E{\propto} e^{-\alpha^{\rm \rm H,NH}(\varepsilon)L}$ (solid lines) properly map the numerical results (markers).  
    (c) The extracted exponent $\alpha^{\rm H,NH}$ as a function of $\varepsilon$. The numerical simulation (markers) is well described by fitting function   
     $\alpha^{\rm H}(\varepsilon){\propto}\ln \varepsilon^{-1.033}$ and $\alpha^{\rm NH}(\varepsilon){\propto}\ln \varepsilon^{-2.089}$, for H-DTC and NH-DTC, respectively.}
    \label{fig:DeltaE}
\end{figure}
{\it Robustness analysis.--}
For $\varepsilon{=}0$, the quasienergy gap between $\pi$-Paired eigenstates $|\psi_{\pm}\rangle$ is exactly $\Delta E{=}|E_{+}{-}E_{-}|{=}\pi$, leading to the perpetual oscillations of the imbalance at frequency $\omega_0{=}\pi$. 
However, $\varepsilon{\neq}0$ in finite-size systems causes a small deviation in the quasienergy gap denoted by $\delta E{=}|\pi{-}\Delta E|$ that results in $|\langle \psi(0) |U_F^{2}|\psi(0)\rangle|^{2}{=}\cos^2(\delta E)$ and 
$|\langle \tilde{\psi}(0) |U_F^{2}|\psi(0)\rangle|^{2}{=}\sin^2(\delta E)$ 
(see SM). 
This shows that the period-doubling evolution lasts for a finite period of time and after timescale $\tau{=}\pi{/}{2\delta E}$ the initial state evolves to its inversion partner. 
Reducing $\delta E$ evidently enhances the stability of DTC by extending its lifetime $\tau$.
This highlights the significance of mechanisms that can effectively result in smaller $\delta E$.
In the following, we show that imperfect incoherent pulses characterized by $(\varepsilon_a,\varepsilon_b){=}(\varepsilon,-\varepsilon)$ can yield significantly smaller $\delta E$'s compared to coherent pulses with $(\varepsilon_a,\varepsilon_b){=}(\varepsilon,\varepsilon)$. 
In Figs.~\ref{fig:DeltaE}(a) and (b) we depict $\delta E$ as a function of $L$ for different values of $\varepsilon$ for the H-DTC and  the NH-DTC, respectively. 
The numerical simulations can be well-fitted  as $\delta E^{\rm H,NH}{\propto}e^{-\alpha^{\rm H,NH}(\varepsilon)L}$, where exponents $\alpha^{\rm H}(\varepsilon),\alpha^{\rm NH}(\varepsilon){>}0$. Here, the superscripts ${\rm H}$ and ${\rm NH}$ refer to H-DTC and NH-DTC, respectively. These results lead to exponential scaling of the timescale $\tau^{\rm H,NH}{\propto}e^{\alpha^{\rm H,NH}(\varepsilon)L}$.
Note that, for $\varepsilon{<}0.2$ in NH-DTC $\delta E$, rapidly diminishes, leading to numerical intractability of the problem. 
In Fig.~\ref{fig:DeltaE} (c), we report the numerically extracted exponents $\alpha^{\rm H}$ and  $\alpha^{\rm NH}$ as a function of $\varepsilon$. The obtained results are well-fitted by $\alpha^{\rm H}(\varepsilon){\propto}\ln \varepsilon^{-1.033}$ and $\alpha^{\rm NH}(\varepsilon){=} \ln \varepsilon^{-2.089}$, indicating that $\alpha^{\rm NH}(\varepsilon){\simeq}2\alpha^{\rm H}(\varepsilon)$. Therefore, the comparison between the life-time of the H-DTC versus NH-DTC shows  
\begin{equation}\label{Eq.Timescale}
 \tau^{\rm NH}\propto \big(\tau^{\rm H}\big)^2.   
\end{equation}
We consider this finding as the key result of this letter, demonstrating that exploiting properly tailored non-reciprocity stabilizes the DTC against pulse imperfections, as evidenced by Eq.~(\ref{Eq.Timescale}). \\

{\it PT-symmetry in NH-DTC.--} The PT-symmetry, i.e. the symmetry under the simultaneous action of parity and time-reversal operators, plays a crucial role in the physics of our NH-DTC. Interestingly, the Hamiltonian Eq.~(\ref{Eq.Hamiltonian}) is indeed PT symmetric, i.e. it commutes with the anti-unitary parity-time operator $\mathcal{PT}$ (see SM), and as a consequence  $[H_{\rm eff},\mathcal{PT}]{=}0$. Owing to the anti-unitary nature of $\mathcal{PT}$, this commutation does not guarantee the existence of a common set of eigenbasis. This indicates the presence of two distinct PT-symmetric phases: an unbroken phase, where $H_{\rm eff}$ and $\mathcal{PT}$ share a common eigenstate basis, and a broken PT-symmetric phase, where such a basis does not exist~\cite{El-ganainy2018}. In the SM, we show that for all values of $(\varepsilon_a,\varepsilon_b)$, the system belongs to an unbroken PT-symmetric phase. As a consequence, the quasi-energy spectra of $H_{\rm eff}$ are real,  namely $E_k{=}E_k^*$, akin to an effective Hermitian Hamiltonian.
Note that real energy spectra are essential for persistent oscillations in Floquet systems, as the imaginary part of complex energy spectra, describing either decaying or amplification processes, would cause either oscillation damping or dynamical instabilities.\\

\begin{figure}
    \centering
\includegraphics[width=0.49\linewidth]{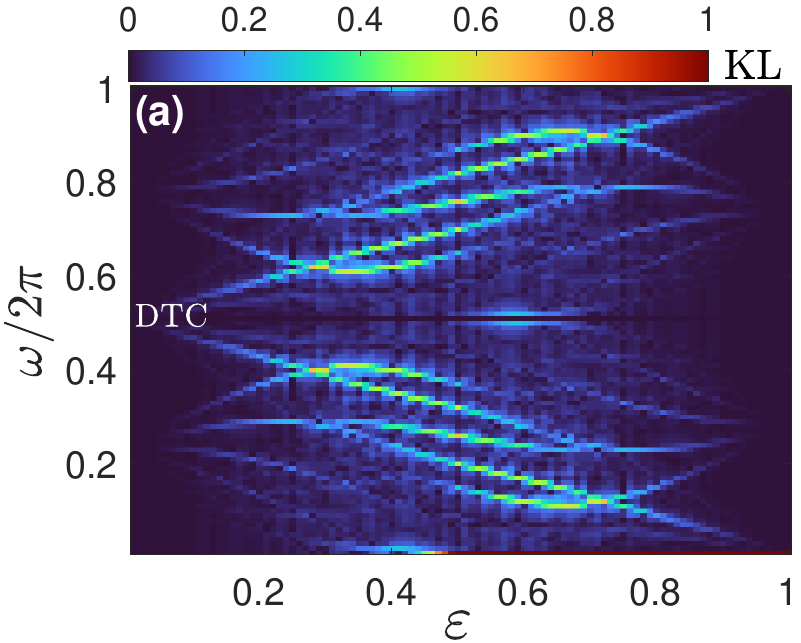}
\includegraphics[width=0.49\linewidth]{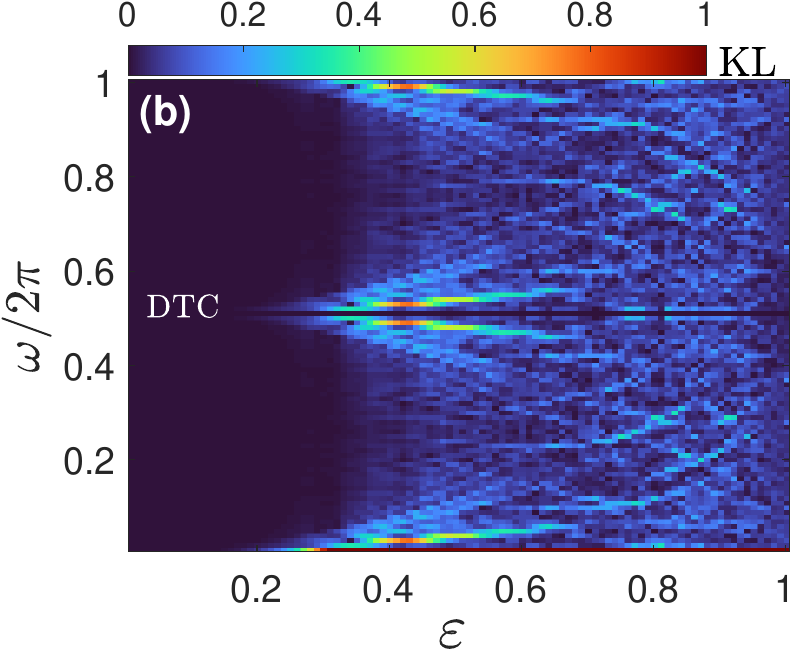}
\includegraphics[width=0.49\linewidth]{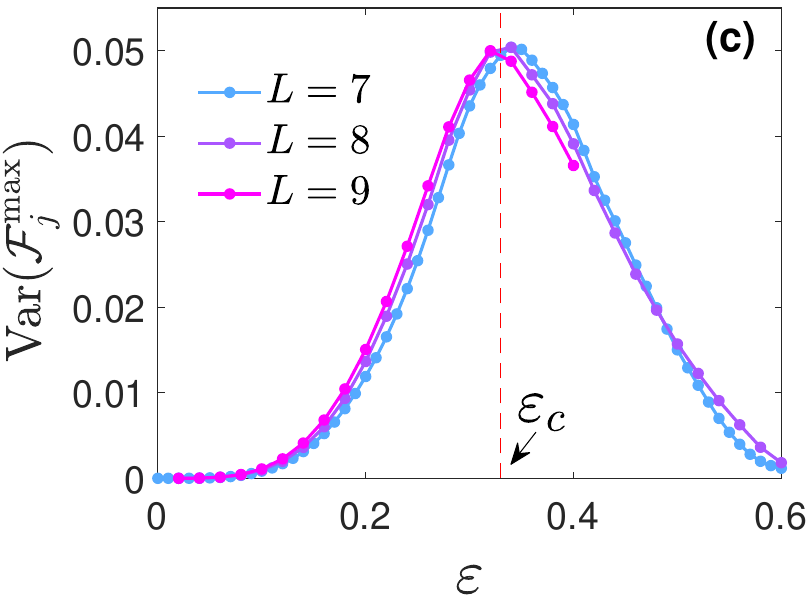}
\includegraphics[width=0.49\linewidth]{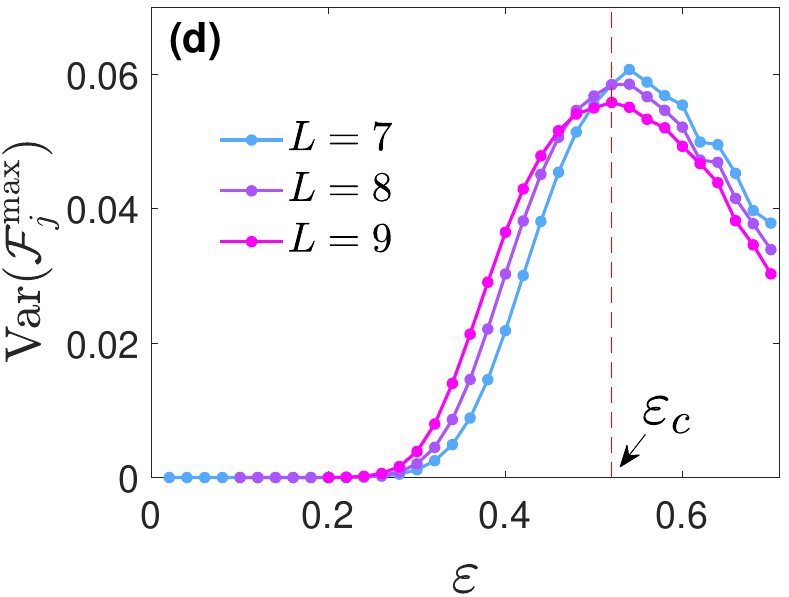}
\caption{The normalized KL divergence as a function of frequency $\omega$ and $\varepsilon$ in (a) H-DTC and (b) NH-DTC with size $L{=}8$, over $n{=}100$. 
The variance of subharmonic Fourier peak height $\mathcal{F}_{j}^{\max}$ as a function of $\varepsilon$ for (c) H-DTC and (d) NH-DTC over $n{=100}$.} \label{fig:KL}
\end{figure}

{\it Melting transition of the DTC.--}
To analyze the melting of the DTC as the imperfection $\varepsilon$ increases, we compute the Fourier spectrum of $\mathcal{I}(nT)$, with the amplitude denoted by $\mathcal{F} (\omega,\varepsilon)$, over $n{=}100$ periods of evolution.  Period-doubling oscillations of $\mathcal{I}(nT)$ result in a subharmonic Fourier peak of $\mathcal{F}(\omega,\varepsilon)$ at $\omega{=}\omega_{0}{=}\pi$. 
To illustrate the modification of Fourier distribution by $\varepsilon$, we use
Kullback-Leibler (KL) divergence, defined as 
\begin{equation}
    \rm{KL}(\omega,\varepsilon) = \mathcal{F}(\omega,\varepsilon) \ln\big(\frac{\mathcal{F}(\omega,\varepsilon)}{\mathcal{F}(\omega,0)}\big).
\end{equation}
In Figs.~\ref{fig:KL}(a) and (b) we depict the KL divergence as a function of $\omega$ and $\varepsilon$ for H-DTC and NH-DTC, respectively. 
In both cases, as we increase the imperfection, $\mathcal{F}(\omega,\varepsilon)$ deviates from $\mathcal{F}(\omega,0)$, which is manifested in the growth of the KL divergence. 
As a result, one observes the transition from the DTC phase for ${\rm KL}{\sim}0$ to the non-DTC regime for ${\rm KL}{\gg}0$. 
Consistent with our earlier observation in Fig.~\ref{fig:Spectrum}, the DTC phase expands to larger values of $\varepsilon$ in the non-reciprocal system considered here.   
To determine the transition point between the DTC phase and non-DTC regime,  following the approach of Ref.~\cite{yao2017discrete}, we
focus on $\mathcal{F}_{j}(\omega,\varepsilon)$ as the Fourier Transform of $\mathcal{I}_{j}(nT)$ for the $j$th site of both chains. 
Analyzing the height of the subharmonic Fourier peaks of $\mathcal{F}_{j}(\omega,\varepsilon)$ at $\omega{=}\omega_0{=}\pi$ for each site $j$, denoted by $\mathcal{F}_j^{\max}$, reveals that near the transition point, one observes strong fluctuations in the values of $\mathcal{F}_j^{\max}$s. 
In Figs.~\ref{fig:KL}(c) and (d), we plot the variance of $\mathcal{F}_j^{\max}$s, namely ${\rm Var}(\mathcal{F}^{\rm max}_{j}){=} \frac{1}{L}\sum_{j{=}1}^{L}(\mathcal{F}^{\rm max}_{j} - \overline{\mathcal{F}^{\rm max}})^2$ where $\overline{\mathcal{F}^{\rm max}}$ represents the average value, as a function of $\varepsilon$ for different system sizes for H-DTC and NH-DTC, respectively. 
Clearly ${\rm Var}(\mathcal{F}_j^{\max})$ for all studied $L$'s closely peaks at some $\varepsilon$, determining the transition point $\varepsilon_{c}$. 
Note that we obtained $\varepsilon_c^{\rm H}{\cong}0.33$ and $\varepsilon_c^{\rm NH}{\cong}0.52$ for H-DTC and NH-DTC, respectively. Indeed, the observation of $\varepsilon_c^{\rm NH}{>}\varepsilon_c^{\rm H}$ further proves that the NH-DTC is more robust against imperfection $\varepsilon$ compared to the H-DTC. 

\begin{figure}
    \centering
    \includegraphics[width=0.49\linewidth]{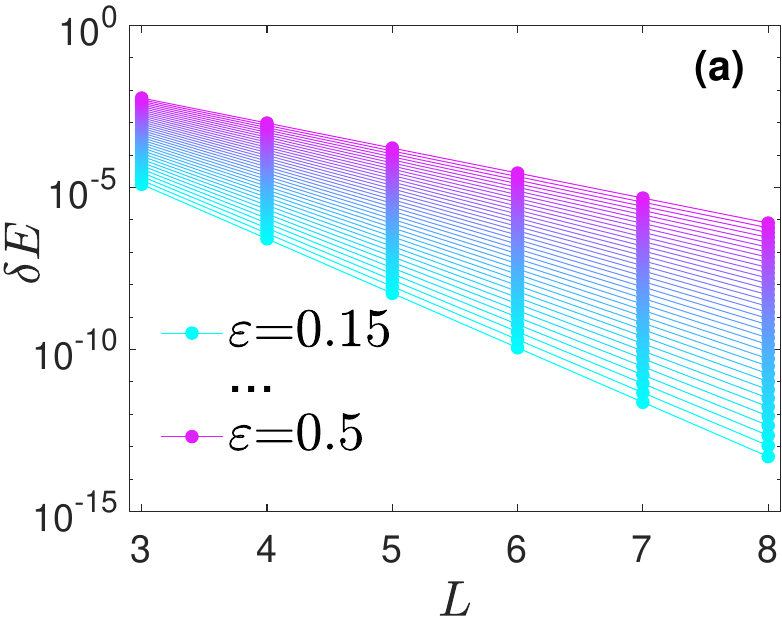}
    \includegraphics[width=0.49\linewidth]{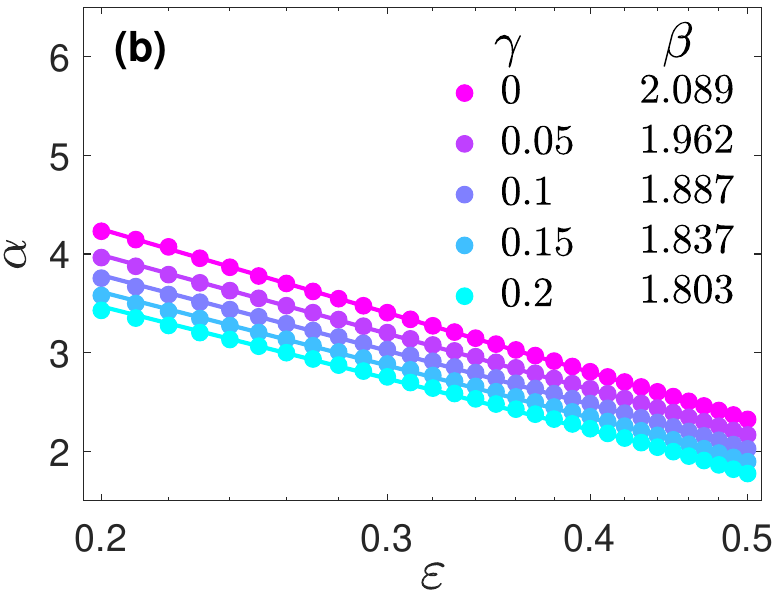}
    \caption{(a) Deviation of the quasienergy gap $\delta E$ as a function of $L$ for $(\varepsilon_a,\varepsilon_b){=}(\varepsilon,-(1{+}\gamma)\varepsilon)$ with $\gamma{=}0.2$ when $\varepsilon$ varies. The numerical results (markers) are properly mapped with the fitting function $\delta E{\propto}e^{-\alpha^{\rm NH}(\gamma,\varepsilon)L}$ (solid lines). (b) The extracted exponent $\alpha^{\rm NH}(\gamma,\varepsilon)$ as a function of $\varepsilon$ for various values of $\gamma$. The numerical results (markers) are well fitted by $\alpha^{\rm NH}(\varepsilon){\propto}\ln \varepsilon^{-\beta(\gamma)}$ (solid lines) for the reported $\gamma$s.}
    \label{fig:GammaEffect}
\end{figure}

{\it Effect of the initial state.--}
Up to now, we focused on the physics of the DTC for a chain initialized in $|\psi(0)\rangle$. 
However, the existence of the DTC in particular for infinitesimal values of $\varepsilon$, in both the Hermitian and non-Hermitian cases, is independent of the initial state. 
Nevertheless, not all spin configurations show the same robustness to the imperfection $\varepsilon$.
In fact, to have a faithful DTC, the initial state has to be a superposition of exactly two $\pi$-paired eigenstates. Those initial states, which have overlap with more than one pair, cannot show stable period-doubling oscillations as time-dependent phases between the different pairs do not cancel. 
In SM, we mapped out the effect of the deviation from $|\psi(0)\rangle$ on the performance of both H-DTC and NH-DTC. 
The results show that, for reasonable values of $\varepsilon$ and over tens period cycles, one has a persistent period-doubling oscillation of 
$\mathcal{I}(nT)$, even when the initial state is different from $|\psi(0)\rangle$.
Furthermore, our results show that the non-reciprocal system considered here leads a DTC with more robustness against variations of the initial state, in comparison with the H-DTC.
\\

{\it NH-DTC with $|\varepsilon_{a}|{\neq}|\varepsilon_{b}|$.--}
Having elucidated the physics of the NH-DTC with $(\varepsilon_{a},\varepsilon_{b}){=}(\varepsilon,-\varepsilon)$, we now explore the more general case, namely $|\varepsilon_{a}|{\neq}|\varepsilon_{b}|$.
In this case, we consider $(\varepsilon_{a},\varepsilon_{b}){=}(\varepsilon,{-}(1{+}\gamma)\varepsilon)$ (with $0{<}\gamma{<}\varepsilon$) and aim to extract the corresponding $\alpha^{\rm NH}(\varepsilon,\gamma)$ using the same method explained previously for $\gamma{=}0$.
In Fig.~\ref{fig:GammaEffect} (a), the deviation of the quasienergy gap, $\delta E^{\rm NH}$, is plotted as a function of $L$ when $\varepsilon$ varies and $\gamma{=}0.2$. The numerical results are well mapped by the fitting function $\delta E^{\rm NH}{\propto}e^{-\alpha^{\rm NH}(\varepsilon,\gamma)L}$.
Fig.~\ref{fig:GammaEffect} (b) depicts the obtained $\alpha^{\rm NH}$'s   as a function of $\varepsilon$ for different values of $\gamma$. Numerical results are well described by the fitting function $\alpha^{\rm NH}(\varepsilon,\gamma){\propto}\ln \varepsilon^{-\beta(\gamma)}$. 
Here, the exponent $\beta$, reported in Fig.~\ref{fig:GammaEffect} (b),  decreases slightly from $\beta(\gamma{=}0){=}2.089$ to $\beta(\gamma{=}0.2){=}1.803$, indicating that the non-reciprocity in the dynamic can considerably stabilize the DTC against imperfection however the optimal stabilization is achieved if $(\varepsilon_{a},\varepsilon_{b}){=}(\varepsilon,-\varepsilon)$.

{\it Conclusion.--} 
An essential criterion for having a faithful discrete time crystal (DTC) phase in an isolated system is to have a certain level of robustness against imperfections in driving pulses.  
In this letter, we present a mechanism to establish a stable DTC in a quantum system, governed by a properly tailored non-Hermitian Hamiltonian.
We show that by engineering non-reciprocity, as the unique characteristic of the non-Hermitian Hamiltonians, in the dynamics of the system one can improve the stability of the DTC against pulse imperfection.
In fact, such enhancements are due to the emergent eigenstate ordering in the spectrum of our non-Hermitian Hamiltonian which prevents the system from thermalization. 
The proposed DTC is achievable by a large class of initial states.

{\it Acknowledgement.--} 
A.B. thanks the National Natural Science Foundation of China (Grants No. 12050410253, No. 92065115, and No. 12274059), and the Ministry of Science and Technology of China (Grant No. QNJ2  021167001L) for their support. K.S. acknowledges the support of the
National Science Centre, Poland, via Project No. 2021/42/A/ST2/00017. A.C. acknowledges support from European Union – Next Generation EU through projects: Eurostart 2022 Topological atom-photon interactions for quantum technologies MUR D.M. 737/2021); PRIN 2022- PNRR no. P202253RLY Harnessing topological phases for quantum technologies; THENCE – Partenariato Esteso NQSTI – PE00000023 – Spoke 2."


\pagebreak
\widetext
\clearpage
\begin{center}
\textbf{\large Supplementary Materials}
\end{center}

\setcounter{section}{0}
\setcounter{equation}{0}
\setcounter{figure}{0}
\setcounter{table}{0}
\setcounter{page}{1}
\makeatletter
\renewcommand{\theequation}{S\arabic{section}}
\renewcommand{\theequation}{S\arabic{equation}}
\renewcommand{\thefigure}{S\arabic{figure}}
\renewcommand{\thetable}{S\arabic{table}}
\renewcommand{\bibnumfmt}[1]{[S#1]}
\renewcommand{\citenumfont}[1]{S#1}

\section{Finite size effect and emergence of $\delta E$}
\begin{figure}[h]
    \centering
    \includegraphics[width=0.24\linewidth]{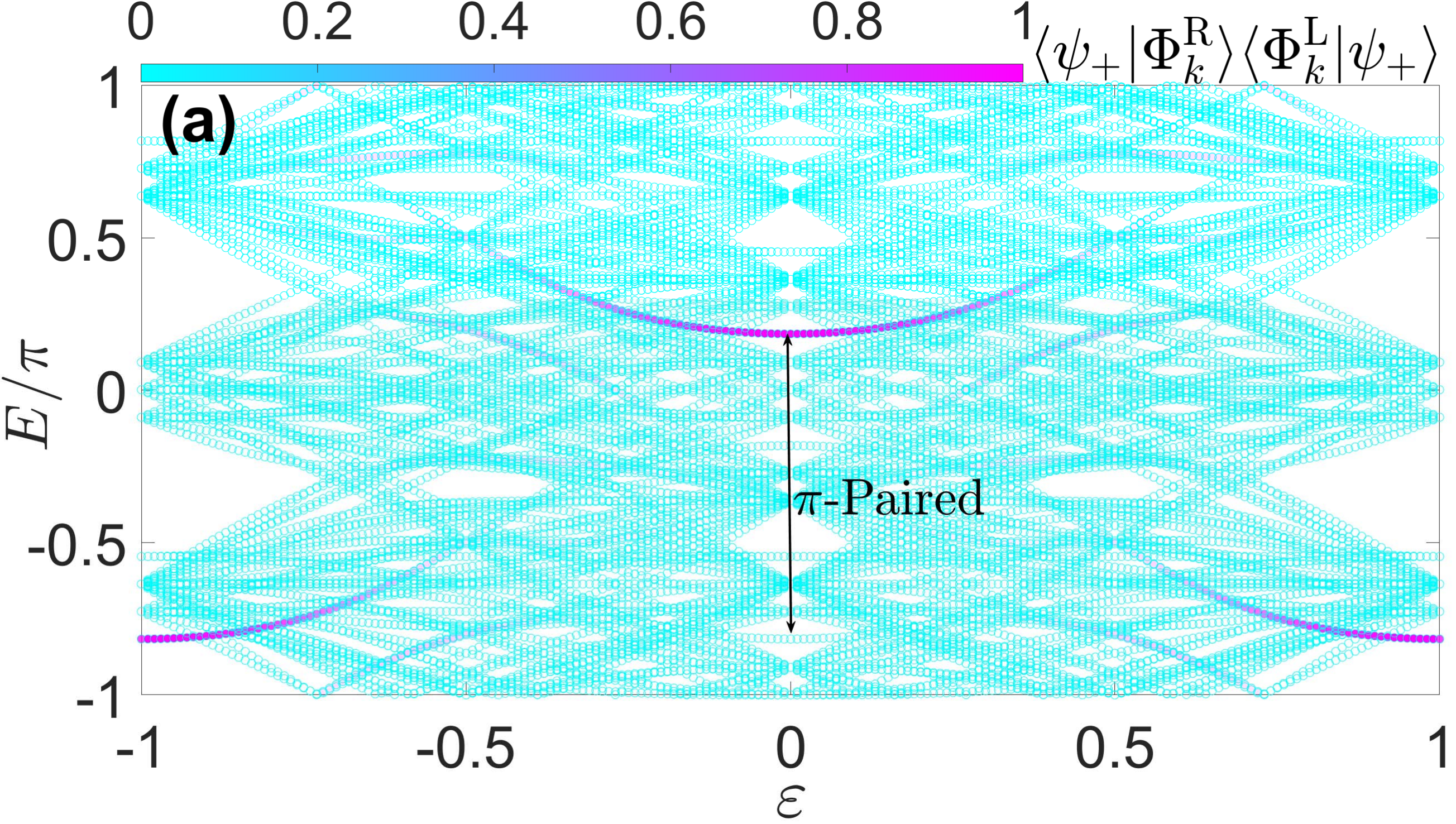}
    \includegraphics[width=0.24\linewidth]{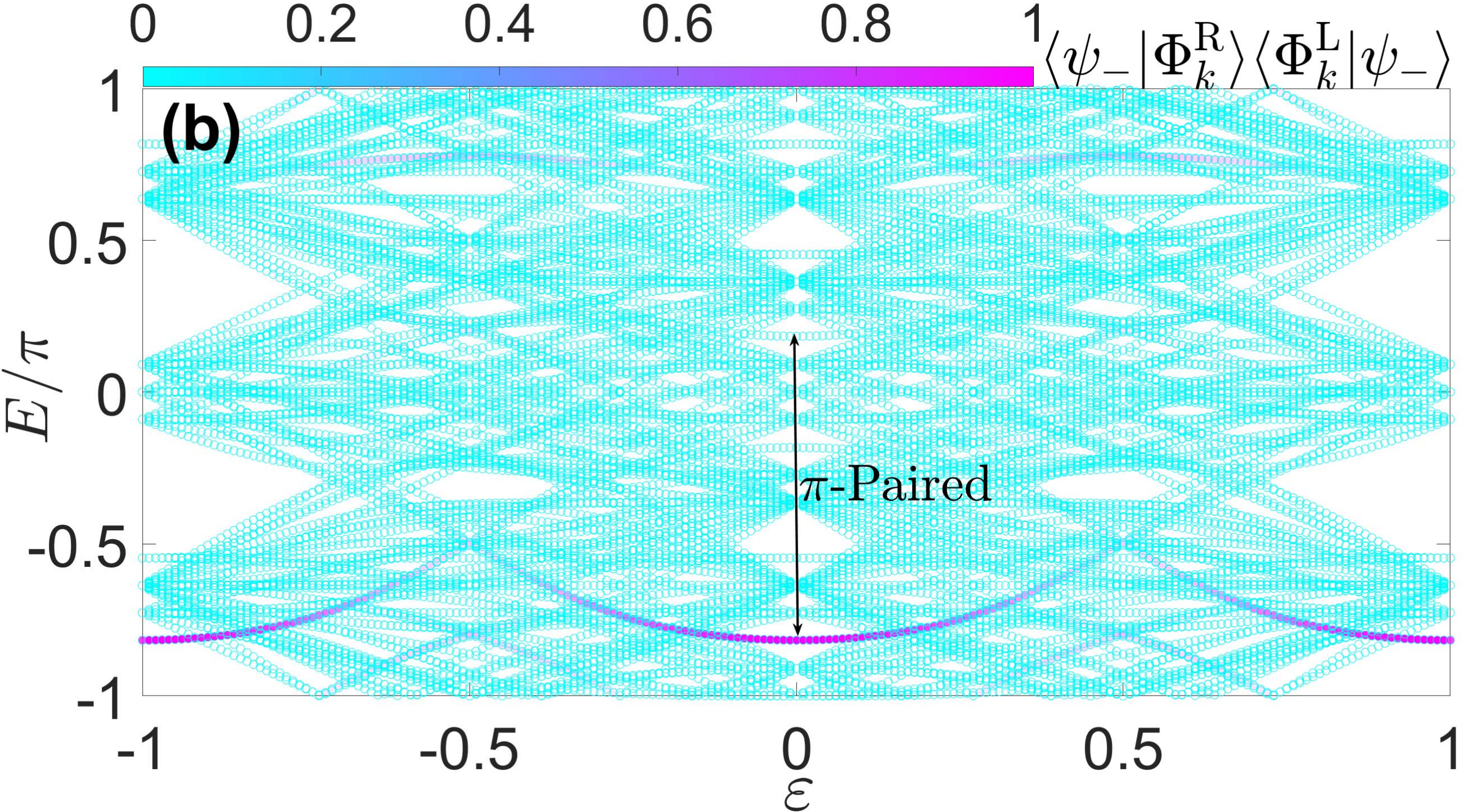}
    \includegraphics[width=0.24\linewidth]{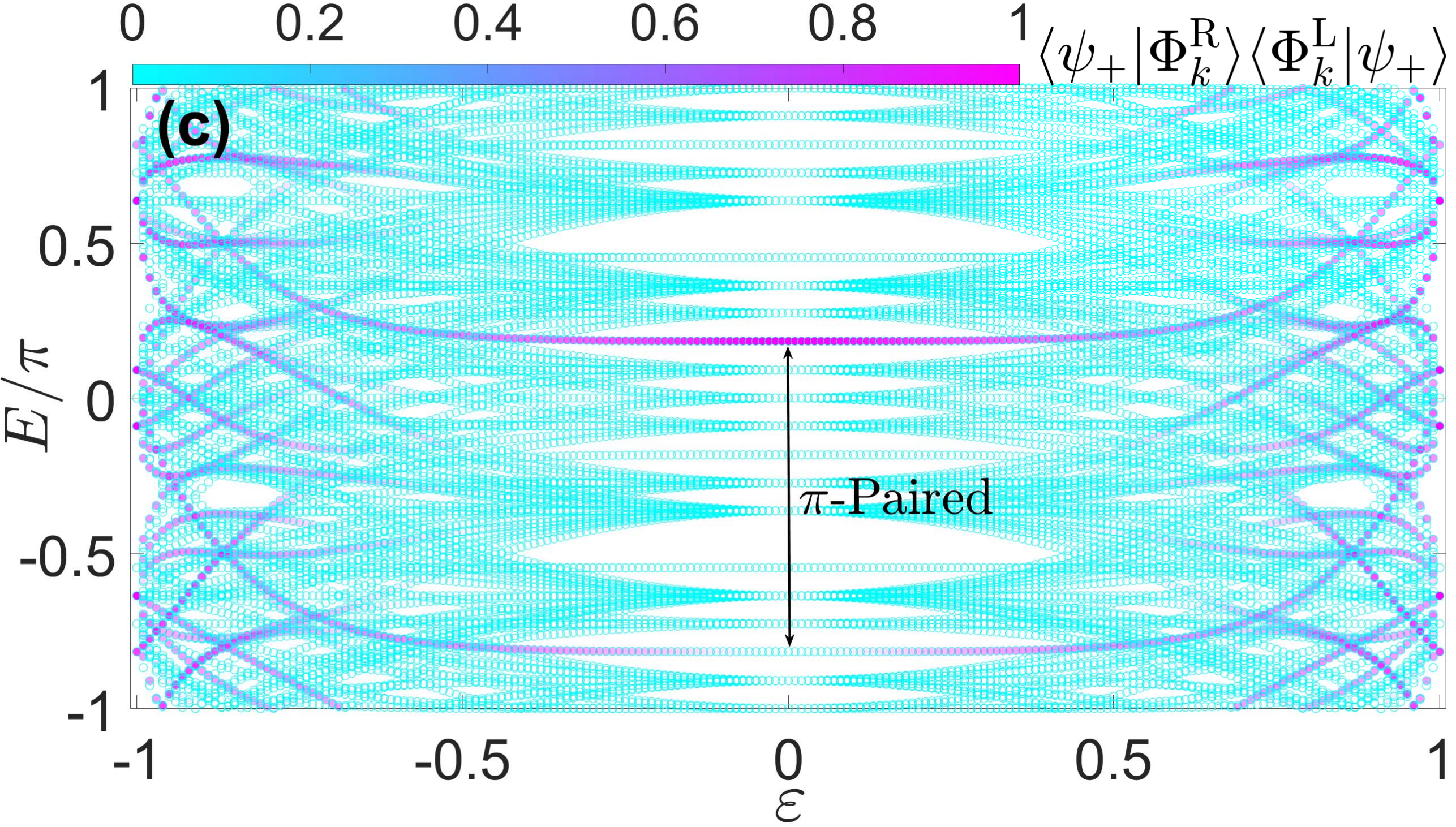}
    \includegraphics[width=0.24\linewidth]{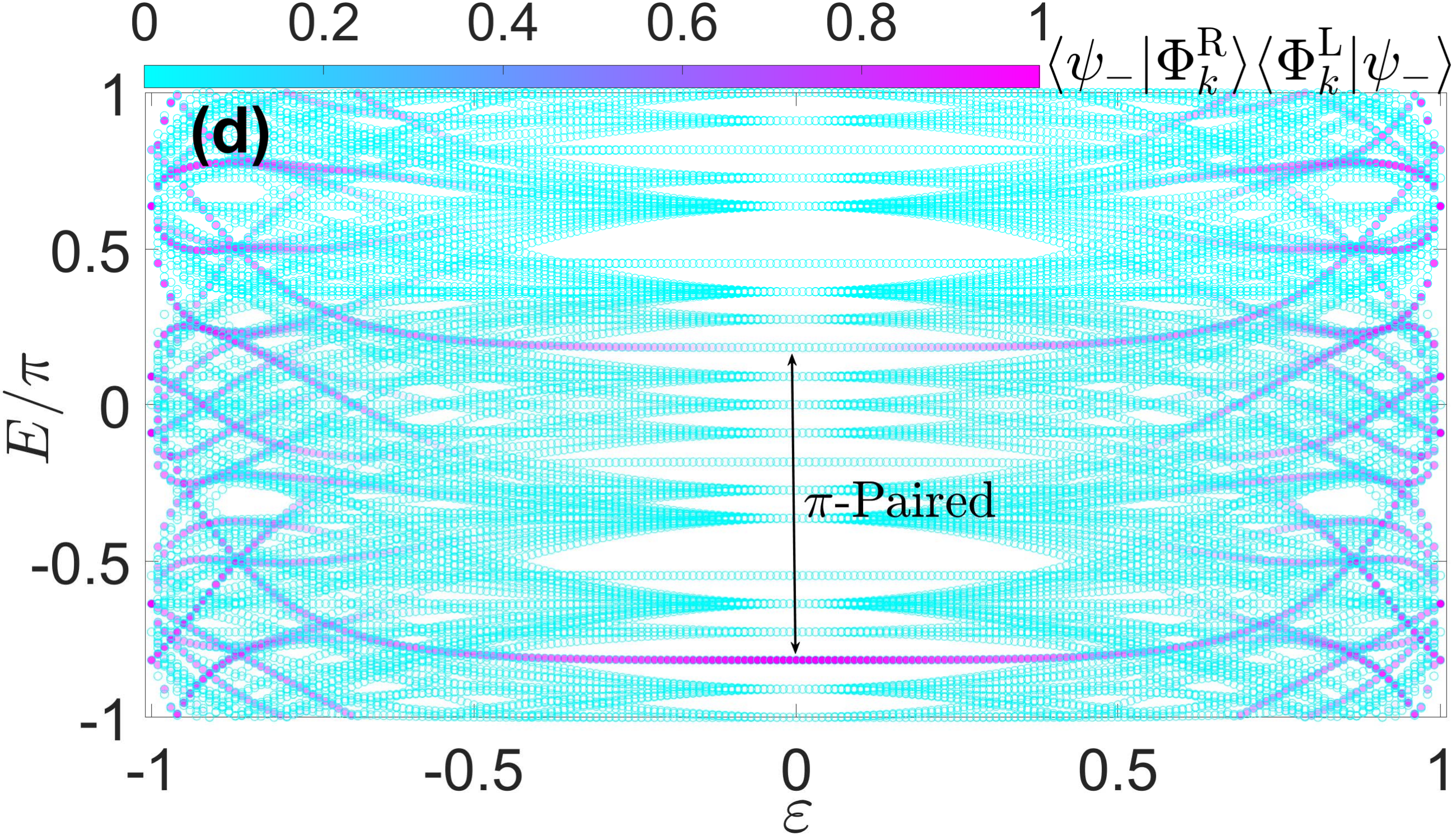}
    \caption{The overlap of the Floquet states with $|\psi_{\pm}\rangle{=}(|\psi(0))\rangle{+}|\tilde{\psi}(0))\rangle)/\sqrt{2}$ as a function of the quasienergy $E$ and the imperfection $\varepsilon$, in (a)-(b) H-DTC, namely  $(\varepsilon_a,\varepsilon_b){=}(\varepsilon,\varepsilon)$, and (c)-(d) NH-DTC, namely $(\varepsilon_a,\varepsilon_b){=}(\varepsilon,-\varepsilon)$, in a system of size $L{=}6$.}
    \label{fig:SM_Spec}
\end{figure}

In the main text, we discuss the appearance of a small deviation in the gap between the quasi-energy of the main eigenstates of $U_{F}$ that have the highest overlap with the initial states.
In this section, we elaborate on this issue.
Starting with an initial state $|\psi(0)\rangle$, one has 
\begin{equation}\label{Eq.S1}
\big|\langle \psi(0) |U_{F}^{2}| \psi(0) \rangle \big|^2 = \Big|\sum_{l,k}
e^{-i(E_l + E_k)}
\langle \psi(0) |\Phi_{l}^{\rm R}\rangle 
\langle \Phi_l^{\rm L}|\Phi_k^{\rm R}\rangle 
\langle \Phi_k^{\rm L}| \psi(0) \rangle \Big|^2  
=
\Big|\sum_{l}
e^{-2iE_l}
\langle \psi(0) |\Phi_{l}^{\rm R}\rangle  
\langle \Phi_l^{\rm L}| \psi(0) \rangle \Big|^2,
\end{equation}
where we used biorthogonality of the Floquet states, namely $\langle \Phi_{l}^{L}|\Phi_{k}^{R} \rangle {=} \delta_{lk}$.
The results presented in Fig.~(2) of the main text 
show that there are dominant $\pi$-Paired eigenstates with corresponding eigenenergy $E_{\pm}$ that show strong overlap with $|\psi(0)\rangle$ and its inversion partner  $|\tilde{\psi}(0)\rangle$. 
In Fig.~\ref{fig:SM_Spec} we plot the overlap of the “Schr{\"o}dinger cat” states as $|\psi_{\pm}\rangle{=}(|\psi(0)\rangle {\pm}|\tilde{\psi}(0)\rangle){/}\sqrt{2}$ with the Floquet states in both (a)-(b) H-DTC, namely when $(\varepsilon_a,\varepsilon_b){=}(\varepsilon,\varepsilon)$, and (c)-(d)  NH-DTC, namely when $(\varepsilon_a,\varepsilon_b){=}(\varepsilon,-\varepsilon)$ as a function of quasienergy $E$ and imperfection $\varepsilon$.
The results confirm that the $\pi$-Paired eigenstates play the main roles in the dynamics starting from the considered initial state and its inversion partner.
Therefore, one simply has 
\begin{equation}\label{Eq.S2}
\big|\langle \psi(0) |U_{F}^{2}| \psi(0) \rangle \big|^2 \cong
\frac{1}{4}\big|e^{-2iE_{+}} + e^{-2iE_{-}} \big|^2  = 
\frac{1}{2}(1 + \cos(2\Delta E)) = \cos^2(\Delta E). 
\end{equation}
Following the same method, one obtains
\begin{equation}\label{Eq.S3}
\big|\langle \tilde{\psi}(0) |U_{F}^{2}| \psi(0) \rangle \big|^2 \cong
\frac{1}{4}\big|e^{-2iE_{+}} - e^{-2iE_{-}} \big|^2  = 
\frac{1}{2}(1 - \cos(2\Delta E)) = \sin^2(\Delta E). 
\end{equation}
In the ideal case, namely $\varepsilon{=}0$, one has $\Delta E{=}\pi$, and hence $\big|\langle \psi(0) |U_{F}^{2}| \psi(0) \rangle \big|^2{=}1$ and 
$\big|\langle \tilde{\psi}(0) |U_{F}^{2}| \psi(0) \rangle \big|^2{=}0$, hinting the perpetual oscillations of the system prepared in the $| \psi(0) \rangle$ or $| \tilde\psi(0) \rangle$ states at the inherent frequency $\omega_0{=}\pi$. 
The finite size effect appears in systems with limited size under the imperfect pulse, namely for $\varepsilon{\neq}0$.
In this case, the quasi-energy gap of the dominant $\pi$-Paired has a small deviation from $\pi$, denoted by $\delta E{=}|\pi - \Delta E|$. 
Therefore the Eqs.~(\ref{Eq.S2}) and~(\ref{Eq.S3}) reduce to 
$\big|\langle \psi(0) |U_{F}^{2}| \psi(0) \rangle \big|^2{=}\cos^{2}(\delta E)$ and $\big|\langle \tilde{\psi(0)} |U_{F}^{2}| \psi(0) \rangle \big|^2{=}\sin^{2}(\delta E)$.
This calculations show that after the timescale $\tau{=}\pi{/}2\delta E$ the initial state $|\psi(0)\rangle$ is transfered to $|\tilde{\psi}(0)\rangle$.
In the main text, we discuss how $\tau$ increases exponentially by the system size. In particular, we systematically show that properly tailored non-reciprocity in the dynamic caused by engineered decoherence can increase the robustness of the DTC to imperfections and enhance the lifetime of the DTC by the power of two.

\section{PT-Symmetry}
Hermiticity of the Hamiltonian is one of the fundamental requirements in quantum theory, which guarantees the conservation of total probability and reality of the energy spectrum.
The PT-symmetric theory was originally intended to extend quantum mechanics to a complex arena, wherein these two conventional physical requirements are relaxed.
In PT-symmetric systems, the combined operations of parity $\mathcal{P}$, representing spatial inversion, and time-reversal $\mathcal{T}$ lead to intriguing properties of the associated Hamiltonian. 
A key characteristic of PT-symmetric Hamiltonian is preserving reality in their energy spectra under certain conditions, akin to Hermitian systems.
In this case, the Hamiltonian of the system is compatible with the antilinear and antiunitary $\mathcal{PT}$ operator. 
To inspect the validity of this condition in our system, one can check the effect of parity and time-reversal operators
on the elements of the Hamiltonian presented in Eq.~(1). 
Considering both position and spin degrees of freedom~\cite{krasnok2021PTSS,mochizuki2016explicitS}, one has

\begin{equation}
 \mathcal{P} =\sum_{j=-L/2}^{L/2} |{-j}\rangle \langle j| ,
 \quad \quad
  \mathcal{T} = \sum_{\mu\in\{a,b\} }  \sum_{j=-L/2}^{L/2} |j\rangle \langle j |\otimes \sigma^{\mu y}_{j} K,
\end{equation}
here $|-j\rangle \langle j |$ swaps the position of spins on sites $j$ and $-j$, when one labels the chain as $\{-L/2,\cdots,-1,0,1,\cdots,L/2\}$. The Pauli operators act on the spin degrees of freedom on $j$th site and $K$ is the antilinear complex conjugation operator, which acts as $K i K {=}{-}i$, with $K^2{=}I$.   
A simple calculation shows that for $\mu{=}a,b$ one has
\begin{equation}
 \mathcal{P}  \mathcal{T} \sigma^{\mu  z}_{j} (\mathcal{T})^{-1} \mathcal{P} = -\sigma^{\mu  z}_{-j},  \quad  \quad
  \mathcal{P}  \mathcal{T} \sigma^{\mu  \pm}_{j} (\mathcal{T})^{-1} \mathcal{P} = -\sigma^{\mu  \pm}_{-j}.
\end{equation}
Substituting these expressions in the Hamiltonian Eq.~(1) shows that $(\mathcal{P}\mathcal{T}) H(t) (\mathcal{P}\mathcal{T})^{-1} {=} H(t)$ and, hence, the Hamiltonian is PT-symmetric. 
As already mentioned in the main text, due to the anti-linear nature of the operator $\mathcal{P}\mathcal{T}$, showing that a given Hamiltonian $H$ commutes with $\mathcal{P}\mathcal{T}$ does not ensure that these two operators share a common eingenbasis.
This points to the identification of two distinct PT-symmetric phases: an unbroken phase in which $H$ and $\mathcal{PT}$ share a common eigenstate basis, and a broken PT-symmetric phase where no such basis exists~\cite{Elganainy2018S}. In the first case, the spectrum is guaranteed to be real, whereas in the latter, the spectrum displays pairs of complex conjugate energies.

One can indeed show that a necessary and sufficient condition for a PT-symmetric system to be in a PT-unbroken phase is the existence of a linear operator $\mathcal{C}$, satisfying $[\mathcal{C},H]{=}0$ and $\mathcal{C}^2{=}I$~\cite{Bender2010S}.
It is easy to verify that $\mathcal{P}$ does commute with the Hamiltonian, and indeed $\mathcal{P}^2{=}I$.
This shows that regardless of the Hamiltonian parameters, the system does not break the PT-symmetry, and its energy spectrum is always real.

\section{Effect of the initial states}
\begin{figure}[h]
    \centering
    \includegraphics[width=0.20\linewidth]{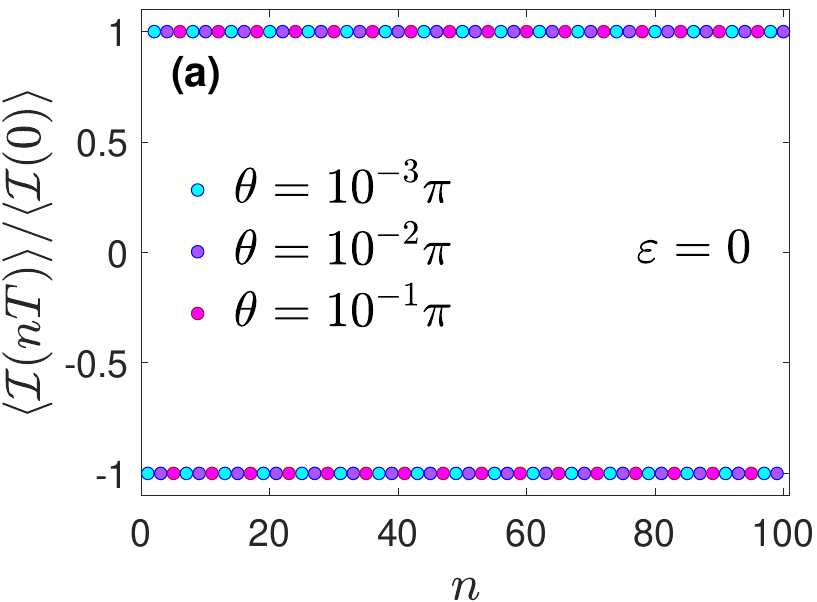}
    \includegraphics[width=0.195\linewidth]{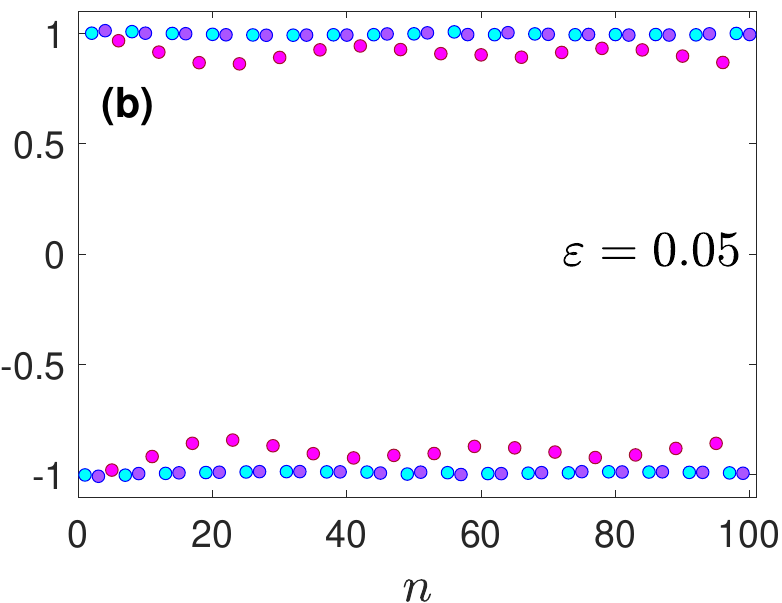}
    \includegraphics[width=0.195\linewidth]{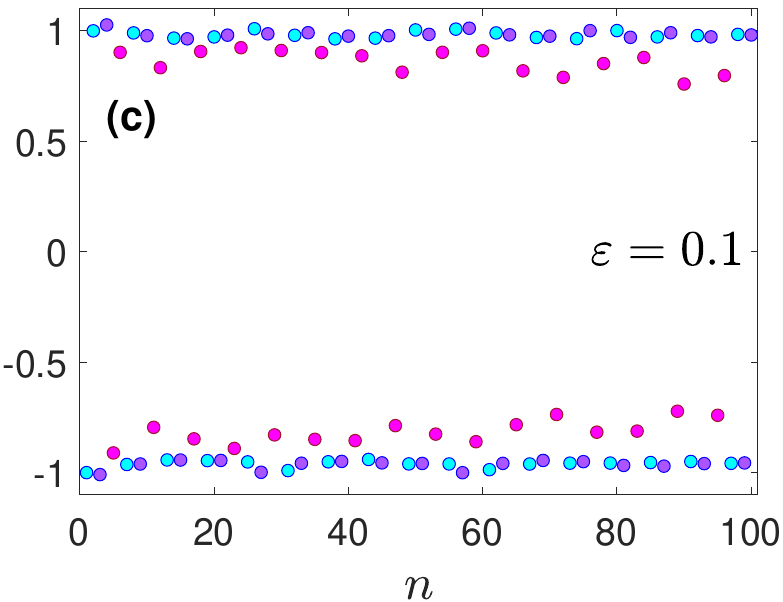}
    \includegraphics[width=0.195\linewidth]{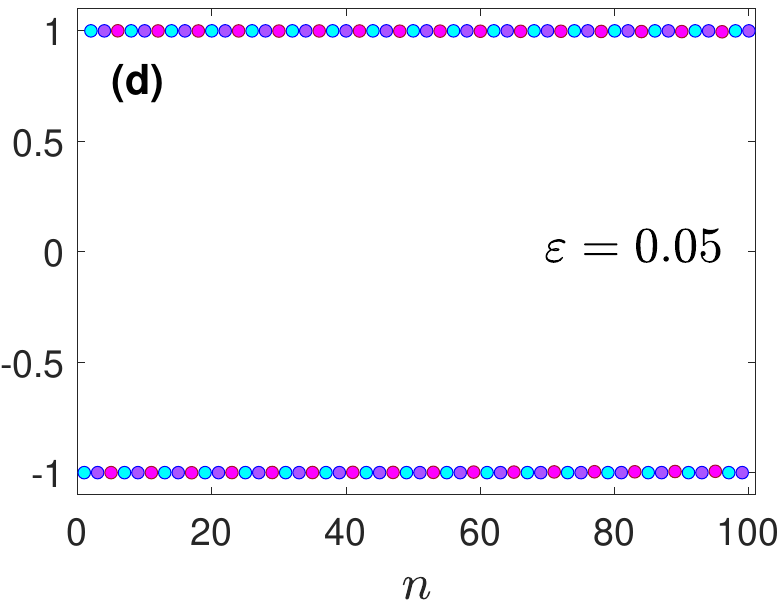}
    \includegraphics[width=0.195\linewidth]{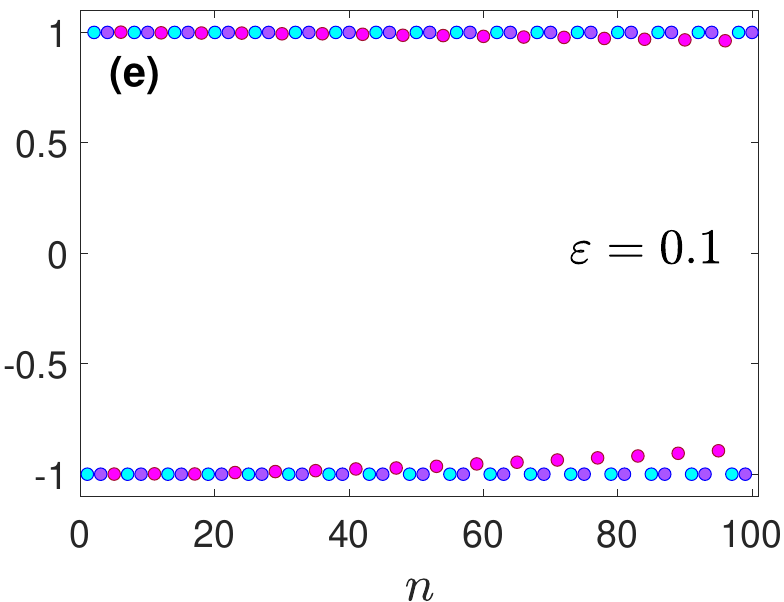}
    \caption{The imbalance in the total magnetization $\mathcal{I}(nT)$ versus period cycling $n$ in (a) perfect DTC, namely $\varepsilon{=}0$, (b)-(c) H-DTC, namely $(\varepsilon_a,\varepsilon_b){=}(\varepsilon,\varepsilon)$, and (d)-(e) NH-DTC, namely $(\varepsilon_a,\varepsilon_b){=}(\varepsilon,-\varepsilon)$. The results are obtained for a system of size $L{=}8$ initialized in Eq.~\ref{Eq.SM_IS}.}
    \label{fig:SM_IS}
\end{figure}

In the main text, we show that initializing the system in state $|\psi(0)\rangle{=}|{\uparrow,\cdots,\uparrow}\rangle_{a}{\otimes}|{\downarrow,\cdots,\downarrow}\rangle_{b}$ provides persistence period-doubling oscillations of $\langle\mathcal{I}(nT)\rangle$ even for $\varepsilon{\neq}0$. 
In general, for $\varepsilon{=}0$, any initial state returns to itself after time $2T$, therefore, period-doubling oscillations of $\langle\mathcal{I}(nT)\rangle$ persists indefinitely even in finite size systems.
For $\varepsilon{>}0$, not all configuration basis states of $H_I$'s support show the same robustness to the imperfection, and
some survive to higher values of $\varepsilon$.
To assess the effect of initial states, we consider a product state as 
\begin{equation}\label{Eq.SM_IS}
|\psi(0)\rangle = \Pi_{j=1}^{L}|\varsigma_j\rangle_{a} \otimes |\varsigma_j\rangle_{b} \quad\quad  |\varsigma_j\rangle_{a} = {\rm cos(\theta)} |\uparrow_j\rangle_{a} +   {\rm sin(\theta)} |\downarrow_j\rangle_{a} \quad\quad   |\varsigma_j\rangle_{b} = -{\rm sin(\theta)} |\uparrow_j\rangle_{b} +   {\rm cos(\theta)} |\downarrow_j\rangle_{b}
\end{equation}
with $\theta{\in}[0,\pi/2]$.
While for the extreme values $\theta{=}0$ and $\theta{=}\pi/2$ one recovers $|\psi(0)\rangle{=}|{\uparrow,\cdots,\uparrow}\rangle_{a}{\otimes}|{\downarrow,\cdots,\downarrow}\rangle_{b}$, and  $|\tilde{\psi}(0)\rangle{=}|{\downarrow,\cdots,\downarrow}\rangle_{a}{\otimes}|{\uparrow,\cdots,\uparrow}\rangle_{b}$, a weighted mixture of different configurations is obtainable for $0{<}\theta{<}\pi/2$.
As here we analyze the imbalance of the magnetization, to hinder mixtures that result in $\langle\mathcal{I}(0)\rangle{=}0$, we focus on $0{<}\theta{<}\pi/4$.
In Fig.~\ref{fig:SM_IS} we present the results for the normalized imbalance $\langle\mathcal{I}(nT)\rangle/\langle\mathcal{I}(0)\rangle$ over tens period cycles when $\theta$ and $\varepsilon$ vary.
While for $\varepsilon{=}0$ and all considered values of $\theta$ one observes perfect oscillations of the imbalance in Fig.~\ref{fig:SM_IS} (a), increasing $\varepsilon$ in both H-DTC, namely $(\varepsilon_a,\varepsilon_b){=}(\varepsilon,\varepsilon)$, and NH-DTC, namely $(\varepsilon_a,\varepsilon_b){=}(\varepsilon,-\varepsilon)$, can affect the oscillations amplitude.
In Fig.~\ref{fig:SM_IS} (b) and (c), the dynamic of imbalance is reported in the H-DTC for two values of the $\varepsilon$.
Analogous results in NH-DTC are plotted in Fig.~\ref{fig:SM_IS} (d) and (e). 
While in H-DTC only infinitesimal deviations from $|\psi(0)\rangle$ can be tolerated,  the non-reciprocity of the dynamic within the NH-DTC increases resilience against these variations.


\end{document}